\documentclass[twocolumn,pre,floatfix]{revtex4}

\usepackage{epsfig}
\usepackage{amsmath}
\usepackage{times}

\begin{document} 

\title{Molecular Dynamics Simulations of Lipid Bilayers: \\
       Major Artifacts due to Truncating 
       Electrostatic Interactions} 

\author{M. Patra and M. Karttunen} 
\affiliation{Biophysics and Statistical Mechanics Group, 
Laboratory of Computational Engineering, Helsinki University 
of Technology, P.O. Box 9203, FIN--02015 HUT, Finland} 

\author{M. T. Hyv\"onen} 
\affiliation{Wihuri Research Institute, Kalliolinnantie 4, 
FIN--00140 Helsinki, Finland, and 
Laboratory of Physics and Helsinki Institute of Physics, 
Helsinki University of Technology, P.O. Box 1100, FIN--02015 HUT, Finland} 

\author{E. Falck, P. Lindqvist, and I. Vattulainen} 
\affiliation{Laboratory of Physics and Helsinki Institute of Physics, 
Helsinki University of Technology, P.O. Box 1100, FIN--02015 HUT, Finland} 

\date{November 27, 2002}

\begin{abstract}
We study the influence of truncating the electrostatic interactions 
in a fully hydrated pure dipalmitoylphosphatidylcholine (DPPC) bilayer 
through 20\,ns molecular dynamics simulations. The computations
in which the electrostatic interactions were truncated are compared 
to similar simulations using the Particle-Mesh Ewald (PME) technique.  
All examined truncation distances (1.8 to 2.5\,nm) lead to major 
effects on the bilayer properties, such as enhanced order of acyl 
chains together with decreased areas per lipid. The results obtained 
using PME, on the other hand, are consistent with experiments. These 
artifacts are interpreted in terms of radial distribution functions 
$g(r)$ of molecules and molecular groups in the bilayer plane. 
Pronounced maxima or minima in $g(r)$ appear exactly at the cutoff 
distance indicating that the truncation gives rise to artificial 
ordering between the polar phosphatidyl and choline groups of the 
DPPC molecules. In systems described using PME, such artificial 
ordering is not present. 
\end{abstract}

\maketitle

\section{Introduction}

One of the great challenges in biophysics is to understand 
the basic principles that govern lipid bilayer mixtures. 
Lipid bilayers, or membranes, govern and mediate various 
biologically relevant processes on the cellular level. 
Transfer of ions through membranes and the function of 
enzymes attached to membranes provide two examples of these  
situations. Besides cellular membranes, lipid membranes are 
present in various man-made applications such as liposomes 
used in novel drug delivery techniques \cite{Lan01} and 
in many natural entities such as lipoproteins. The variety 
of situations where lipid bilayers play an important role is 
truly fascinating and is discussed in a number of review 
articles~\cite{Blo91,Lip95,Mer96,Nag00,Kat01}.

The characteristics of membranes have been extensively  
investigated for many decades, and experiments have provided 
substantial information about the intriguing physicochemical 
aspects of membrane systems \cite{Blo91,Lip95,Mer96,Nag00,Kat01}. 
However, while the experimental approach is the cornerstone 
of membrane research, it is often difficult or even impossible 
to obtain a thorough understanding of the phenomena taking 
place in lipid bilayers by experiments only. Therefore, 
atomistic computer simulation techniques such as classical 
molecular dynamics (MD) have become a standard tool for 
studies of biomembrane systems at the molecular level
\cite{Mer96,Tie97,Fel00,Sai02}.

One drawback of the computational approach is that its success 
depends on various methodological issues such as force fields, 
constraints, and the accuracy of integration schemes for the 
equations of motion \cite{Tie97,Gun98,Chi00,Bes00,Vat02}. In 
particular, the treatment of electrostatic interactions deserves 
special attention, since biomembrane systems are highly charged: 
lipid molecules are either polar or charged and they interact with 
each other, the polar water environment, counterions \cite{Pan02}, 
proteins \cite{Ibr98}, and DNA \cite{Ban99}. Proper treatment 
of electrostatic interactions in MD simulations is therefore one 
of the most important issues in this field and it continues to 
pose significant challenges for computer simulations.

The calculation of electrostatic interactions is typically based 
on solving the Poisson equation for the electrostatic potential 
such that all charged particles and their periodic images are 
taken into account in some systematic fashion. The Ewald summation 
method, its variants~\cite{Sag99} and the fast multipole method 
\cite{Gre87,Fre02} are commonly used techniques that exploit 
this idea. In particular, the Particle-Mesh Ewald (PME) technique 
has been used increasingly often in lipid bilayer simulations 
\cite{Ven00,Sai01,Fel01,Tob01,Pan02}.

Alternatively, one can neglect the long-range Coulombic tail 
and truncate the interactions at some suitable distance, a typical 
choice being 1.5\,--\,2.0\,nm. This technique leads to considerable 
savings in the computational load and hence is widely used. Due to 
the speed-up, it is particularly useful in studies of large systems 
over long times \cite{Lin00,Mar01,Chi02}, and when the computational 
requirements are demanding due to, e.g. long time scales associated 
with complex processes such as membrane fusion \cite{Mar02}. 
The discontinuities in the potential and forces at the cutoff 
radius are typically not considered to be a major issue, since 
they can be handled using various shifting and switching 
techniques \cite{Ste94,Lea01}.

One might expect the artifacts due to truncation, if any, to 
become smaller as the cutoff distance $r_{\rm cut}$ is increased, 
and that for reasonably large cutoffs the system should not be 
influenced by truncation. In practice, however, this is not the 
case. The classical example is water: its bulk properties 
\cite{Alp89,Fel96} and properties at the surfaces of lipid 
monolayers have been found to be affected by truncation 
\cite{Alp93a,Alp93b,Fel96}. Other cases where direct effects 
due to truncation have been observed include peptides, proteins, 
and DNA \cite{Smi91,Sch92,Yor93,Yor95,Ibr98,Nor00}.

Given these findings, it is rather surprising that only Venable 
et al.\ have considered the effects of truncation in the context 
of lipid bilayers \cite{Ven00}. They compared the areas per molecule 
in a DPPC bilayer in the gel phase using systems in which the 
electrostatic interactions had been treated using PME and 
a truncation of 1.2\,nm. They found the results to differ by 
about 4\,\%. To the best of our knowledge, further systematic 
studies of truncation effects in lipid bilayers have not been 
reported. The lack of information is particularly striking in the 
case of the liquid-crystalline ($L_{\alpha}$) phase, which is highly 
relevant from a physiological point of view. Instead, the general 
impression seems to be that truncation may lead to artifacts, but 
they are minor, or even negligible, if the cutoff is longer 
than about 1.8\,nm\, \cite{Alp93a,Alp93b,Fel96,Jak96}.

In this article, we show through an extensive set of 20\,ns 
MD simulations for a fully hydrated pure lipid bilayer of 128 
dipalmitoylphosphatidylcholine (DPPC) molecules in the 
liquid-crystalline phase that the truncation of electrostatic 
interactions can have significant consequences on the properties of 
lipid bilayer systems. We consider several truncation distances from 
1.8 to 2.5\,nm\, and compare them to a case where the Particle-Mesh 
Ewald technique has been applied. We find that the simulations 
where PME has been used lead to an area per lipid molecule 
consistent with experiments, 
while the truncation of electrostatic interactions leads to 
5\,--\,14\,\% smaller values. This dramatic result is reflected 
in various properties of the lipid bilayer, including the probability 
distribution of the area per lipid, the density profile across the 
membrane, and the ordering of acyl chains. In addition to these, 
truncation leads to prominent artifacts in the electrostatic 
potential across the bilayer. We interpret the artifacts in terms 
of radial distribution functions $g(r)$ of molecules and molecular 
groups in the plane of the membrane. The radial distribution 
functions reveal without doubt that truncation leads to 
artificial ordering in the head groups of lipid molecules.

We conclude that the truncation of electrostatic interactions 
may lead to profound artifacts in the properties of lipid bilayer 
systems, and should be used with great care, if at all.

\section{System}

\subsection{ Model and simulation details}

\begin{figure}
\centering
\epsfig{file=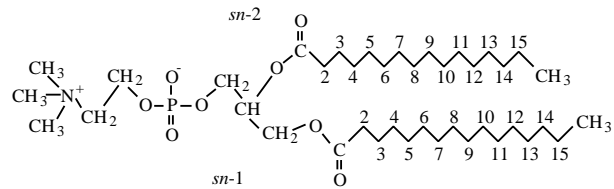,width=3.2in}
~\vspace{0.5cm}
\caption{Representation of a DPPC molecule showing the 
         numbering of carbons in the acyl chains discussed 
         in the text.} 
\label{figDPPCmolecule}
\end{figure}

We have simulated a lipid bilayer system consisting of 128 DPPC 
molecules (shown schematically in Fig.~\ref{figDPPCmolecule}),
fully hydrated by 3655 water molecules. The united atom model 
description used in this work has been validated previously 
\cite{Tie96}.

We used a system available 
at {\tt http://moose.\-bio.\-ucalgary.\-ca/\-files/\-dppc128.pdb}
as our initial configuration. This system corresponds to the 
final structure of run E discussed elsewhere \cite{Tie96}. 
The bilayer is in the $x$-$y$ plane.

The parameters for bonded and non-bonded interactions were
taken from a rather recent study on a DPPC bilayer system 
\cite{Berger97}, available in electronic form at 
{\tt http://moose.bio.ucalgary.ca/files/lipid.itp}. The 
partial charges were obtained from the underlying model 
description \cite{Tie96} and can be found at 
{\tt http://moose.bio.ucalgary.ca/files/dppc.itp}.  
For water, the SPC model \cite{Ber81} was used.

Lennard-Jones interactions were cut off at 1.0\,nm without shift 
or switch functions. Electrostatic interactions within 1.0\,nm 
were calculated at each time step, while interactions beyond 
this range were determined every ten timesteps. These choices 
follow the parameterization of DPPC \cite{Tie96}. 
Long-range electrostatics was handled either by using a cutoff 
at $r_{\rm cut} = 1.8$\,nm, 2.0\,nm, or 2.5\,nm, or by means of 
the Particle-Mesh Ewald \cite{Ess95a} method to take the long-range 
interaction fully into account. The time step for the simulations 
was chosen to be 2.0\,fs.

The simulations were performed using the Gromacs~\cite{Lin01} 
package in the $Np\,T$ ensemble. The Berendsen algorithm with 
a time constant of 1\,ps for pressure coupling was used as barostat. 
The setup was chosen such that the height of the simulation box 
(i.\,e., its extension in the $z$-direction) was allowed to vary 
independently of the cross-sectional area of the box in the 
$x$-$y$ plane. The DPPC and water molecules were separately 
coupled to a heat bath at a temperature $T = 323$\,K using the 
Berendsen algorithm \cite{Ber84} with a coupling constant of 
0.1\,ps. The lengths of all bonds were kept constant with the 
Lincs algorithm~\cite{Hes97}.

The main focus of this paper is on the effects due to different 
treatments of the long-range electrostatic interactions. To this 
end, we have studied DPPC bilayers over a time scale of 20\,ns 
using three different truncation distances and PME. The simulations 
have been repeated at two different temperatures in the 
liquid-crystalline phase to confirm the validity of our conclusions. 
In addition, as described in Appendix~\ref{secAppendix}, we 
performed additional simulations to examine the effects due to 
constraints, time constants of the thermostat and pressure coupling, 
and the range of van der Waals interactions. These simulations 
sum up to about 20 simulations of 20\,ns each. In total, the 
simulations took about 15\,000 CPU hours.

\subsection{Data analysis}
\label{secDataAnalysis}

To calculate the area occupied by each individual lipid and to 
determine the probability distributions for the area per lipid 
$P(A)$, we applied Voronoi analysis in two dimensions 
\cite{Shi98a}. In Voronoi tessellation, we first computed the 
centers of mass (CM) for the lipids and projected them onto the 
$x$-$y$ plane. Thus, the centers of mass define a set of points 
in the $x$-$y$ plane. A point in the plane is considered to belong 
to a particular Voronoi cell if it is closer to the projected CM 
of the lipid molecule associated with that cell than to any 
other CM position.

The mass density profile across the bilayer was calculated 
by separately analyzing each frame of the simulations. The center 
of the bilayer (i.\,e., its $z$-component) was first determined by 
computing the centers of mass for the two monolayers. The positions 
of all atoms were then taken into account with respect to the center. 
It is important to note that the masses of all hydrogen atoms must 
be included explicitly, as has been done in this work. Since the 
system possesses mirror symmetry, all positions with $z < 0$ have 
been folded to $z > 0$ to reduce statistical error.

The electrostatic potential across the bilayer was calculated
in a similar fashion. The average charge density profile was first 
computed in such a way that the center of the bilayer ($z = 0$) 
was determined for each simulation frame separately. Finally, the 
electrostatic potential was determined by integrating the charge 
density twice starting from the initial condition $V(z = 0) = 0$.

The microscopic structure of lipid molecules and the ordering of 
acyl chains is characterized through the order parameter tensor 
$S_{\alpha\beta}$ \, ($\alpha,\beta = x,y,z$) defined as 
\begin{equation}
        S_{\alpha\beta} = 
           \frac{1}{2} 
           \left\langle 
              3 \cos\theta_{\alpha}\,\cos\theta_{\beta} - 
           \delta_{\alpha\beta} 
           \right\rangle 
        \label{sOrderDef},
\end{equation}
where $\theta_{\alpha}$ is the angle between the $\alpha^{\rm th}$  
molecular axis and the bilayer normal ($z$-axis). The order 
parameter is calculated separately for all positions (carbons) 
along the chain. Given the geometry of the bilayer, 
the relevant order parameter is the diagonal element $S_{zz}$. 
This is related to the deuterium order parameter $S_{\mathrm{CD}}$ 
defined as
\begin{equation}
        S_{\mathrm{CD}} = \frac{2}{3} S_{xx} + 
                          \frac{1}{3} S_{yy} \;, 
        \label{sOrderDefCD}
\end{equation}
which is often determined in experiments, e.g.\ using nuclear 
magnetic resonance spectroscopy. Since the bilayer is symmetric 
with respect to rotation around the $z$-axis, we have 
$S_{xx} = S_{yy}$, and $S_{xx} + S_{yy} + S_{zz} = 0$.
Hence, it follows that $S_{\mathrm{CD}} = - S_{zz} / 2$. 
To allow comparison with experimental data, we 
present our results in terms of $| S_{\mathrm{CD}} |$.

Ordering of water in the vicinity of the bilayer-water 
interface is described by calculating the time averaged 
projection of the water dipole unit vector $\vec{\mu}(z)$  
onto the interfacial normal $\vec{n}$,
\begin{equation}
P(z) = \langle \vec{\mu}(z) \cdot \vec{n} \rangle 
     = \langle \cos \theta \, \rangle \, , 
\end{equation}
where $z$ is the $z$-component of the center of mass of the 
water molecule and vector $\vec{n}$ points away from the 
bilayer center along the $z$-coordinate.

In order to calculate the radial distribution functions (RDFs) 
between different charged groups, one should note that the 
groups have internal structure. The positively charged group 
is choline, essentially ${\rm N}({\rm CH}_3)_{3}^{+}$, and 
the negatively charged one is the phosphate group, essentially 
${\rm P}{\rm O}_2{\rm O}^{-}$. To demonstrate possible 
artifacts due to truncation in the pair correlation behavior 
between these charged groups, we found that the most transparent 
way to this end is to consider RDFs between nitrogen and 
phosphate atoms in the choline and phosphate groups. The 
RDFs presented in this work are thus for the P\,--\,P and 
N\,--\,N pairs.

Note that our simulations---like virtually all 
simulations---use a group based cutoff, i.\,e., electrostatic 
interactions are computed for a pair of particles if any pair 
belonging to the two groups is within the cutoff distance. 
Due to this, the system cannot force any atom to a certain 
distance such that it would artificially enhance favorable 
interactions within a group only. As a consequence, the 
artifacts observed using truncated electrostatics cannot 
be explained by the internal structure.

For the radial distribution function between the center 
of mass positions of the DPPC molecules, we first calculated 
the CM for all of them. Then, the $g_{\rm 2d}(r)$ was computed in 
a plane, i.\,e.,\ using the $x,y$\,-coordinates of the CM positions.

\section{Results and discussion}

\subsection{System dimensions}
\label{secGeometry}

\begin{figure}
\centering
\epsfig{file=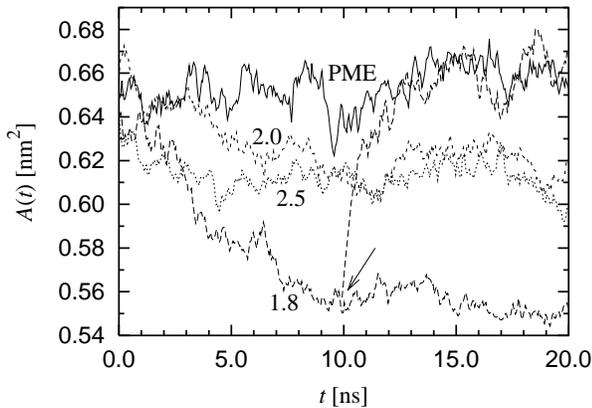,width=3.2in}
~\vspace{0.5cm}
\caption{The area per lipid $A(t)$ as a function of 
         time for truncated and PME electrostatics.
         Cutoff radii are shown in the plot. In addition,
         $A(t)$ is shown for the case where the electrostatics 
         was switched from a cutoff at 1.8\,nm to  PME at 
         10\,ns  (marked by an arrow).}
\label{figAreaTemporal}
\end{figure}

One of the central quantities in describing lipid bilayers 
is the average area per molecule $\langle A \, \rangle$. 
For DPPC it has been experimentally determined to be 
$\langle A \, \rangle = 0.64$\,nm$^2$ \cite{Nag00}.
It provides a measure for tuning the force fields 
and other parameters for lipid systems as one aims for a 
quantitative description of lipid bilayers through MD simulations. 
In the present work, the average area per lipid was computed 
using the size of the simulation 
box in the $x$-$y$ plane. Since we employ pressure coupling, 
the simulation box is allowed to fluctuate during the
simulation. The temporal behavior of the area per lipid $A(t)$ 
is shown in Fig.~\ref{figAreaTemporal}. 
The simulations have equilibrated after 10\,ns
and for the analysis we discarded the first 10\,ns, and 
used the second 10\,ns period only.

Fig.~\ref{figAreaTemporal} shows that $A(t)$ depends strongly 
on the treatment of electrostatic interactions. The simulations 
using PME yielded $\langle A \, \rangle = (0.645 \pm 0.010)$\,nm$^2$ 
consistent with recent experiments \cite{Nag00}. Truncation at 
both 2.0\,nm and 2.5\,nm lead to 
$\langle A \, \rangle = (0.615 \pm 0.010)$\,nm$^2$, which 
deviates about 5\% from the PME result. Further decrease of the 
cutoff distance to 1.8\,nm lead to 
$\langle A \, \rangle = (0.555 \pm 0.010)$\,nm$^2$.
This is about 14\% smaller than the reference value 
of 0.645\,nm$^2$.

We verified that the large differences in the average area per 
lipid are indeed due to electrostatics and not a consequence of
initial conditions. To this end, we chose an equilibrated configuration 
from the simulation with $r_{\rm cut} = 1.8$\,nm (at 10\,ns) as the 
initial configuration for a new simulation. In this new simulation,
the electrostatic interactions were computed using PME instead of
a cutoff. This transition point is marked by an arrow in 
Fig.~\ref{figAreaTemporal}. As seen from Fig.~\ref{figAreaTemporal},
the area per lipid quickly approaches the value of 0.645\,nm$^2$.
We can thus conclude that the different areas per lipid reported 
here are solely due to the treatment of electrostatic interactions.

For the purpose of comparison, let us note that our results for 
the average area per lipid at short times are in agreement with 
earlier \, studies \, by \, Tieleman \, et al.

\begin{figure}
\centering
\epsfig{file=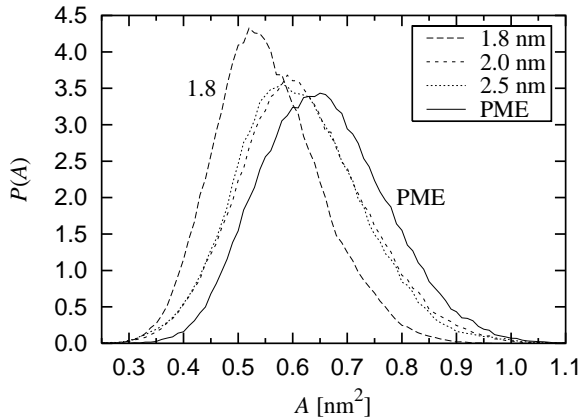,width=3.2in}
~\vspace{0.5cm}
\caption{Distribution of the area per lipid, $P(A)$, computed by 
         the Voronoi analysis.}
\label{figAreaDistribution}
\end{figure}

\noindent
\cite{Tie96}, whose model 
description we follow in the present work. They used a cutoff 
at 2.0\,nm and found $\langle A \, \rangle \approx 0.60$\,nm$^2$ 
for the area per molecule. A full comparison is not meaningful, 
however, since the time scale in their studies was 0.5\,ns and 
the analysis was done over the last 100\,ps only.

We have extended the above studies by considering the probability 
distributions for the area per lipid molecule $P(A)$. This quantity 
is of interest for a number of processes in lipid bilayers, 
e.g.\ the lateral diffusion of lipids in the bilayer plane. 
Results for $P(A)$ are shown in Fig.~\ref{figAreaDistribution}. 
The distributions reveal that the minimum area per lipid is 
approximately $0.3$\,nm$^2$. Further, we find that the shapes 
of the distributions are similar and scaled by the average area 
per molecule. However, it is worth pointing out that even if the 
cutoff distance is increased to a value close to the maximal one 
(i.\,e., half of the linear dimension of the system), the artifacts 
in $P(A)$ persist. This indicates that cutoff distances even as 
large as 2.5\,nm are not sufficient for a proper quantitative 
treatment of electrostatics.

\begin{figure}
\centering
\epsfig{file=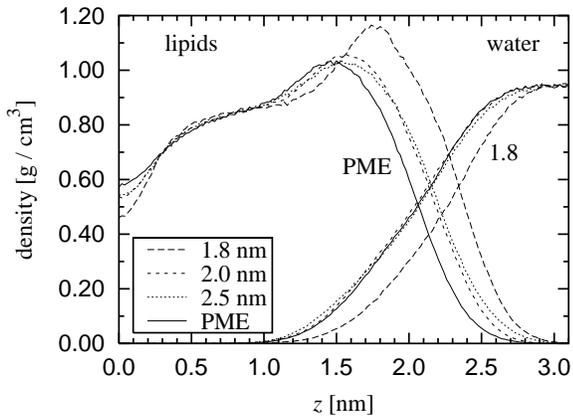,width=3.2in}
~\vspace{0.5cm}
\caption{Mass density profile across the bilayer computed 
         separately for lipids (left) and water (right). $z = 0$
         corresponds to the center of the bilayer and both 
         sides of the bilayer have been folded towards 
         positive $z$.}
\label{figDensityBilayer}
\end{figure}

Next we estimate the density distribution and 
the dimensions along the bilayer normal. The mass density 
profile for water and lipid molecules is shown in 
Fig.~\ref{figDensityBilayer}. As before, we find 
that the smallest cutoff leads to major artifacts as compared 
to the PME description, while the other two truncation 
distances perform better. The results obtained using any truncation 
distance are distinctly different from those obtained using PME, 
however. To demonstrate this, we consider the thickness of the 
bilayer defined as the point where the densities of water and 
lipids are equal. The half-thicknesses of the bilayers were 
found to be 2.03\,nm for PME, 2.15\,nm for $r_{\rm cut} = 2.0$\,nm 
and 2.5\,nm, and 2.33\,nm for $r_{\rm cut} = 1.8$\,nm. It is 
noteworthy that the deviations in these results are of similar 
size as the deviations found in the average area per molecule. 
However, while truncation leads to a {\it decreased} area per 
lipid, it also leads to an {\it increased} height of the membrane. 
Thus these two artifacts compensate each other, and one finds 
that the average volume per lipid obtained by truncation methods 
is consistent with the value found by PME. This result was 
confirmed by the Voronoi analysis for the lipid bilayers in 
three dimensions (data not shown).

It should be noted that the density of the bulk water phase 
is essentially independent of the simulation parameters. This is 
in contrast to the density of the lipid phase. These observations 
confirm that the differences between the simulations are caused 
by the electrostatic interactions between the lipids and/or lipids 
and water, but not by the interactions among the water molecules. 

\bigskip

\subsection{Order parameters}
\label{secLipidOrderParameters}

\begin{figure}
\centering

\epsfig{file=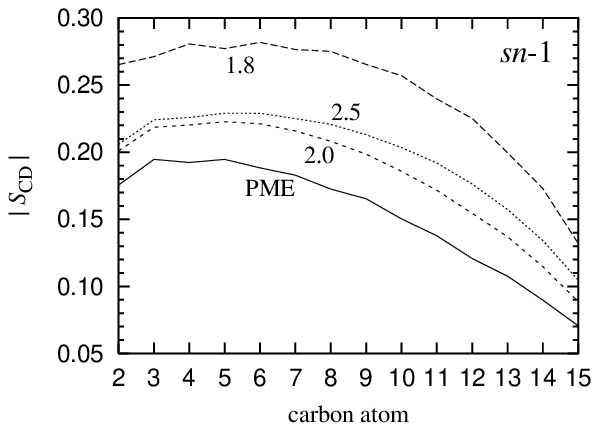,width=3.2in}

\epsfig{file=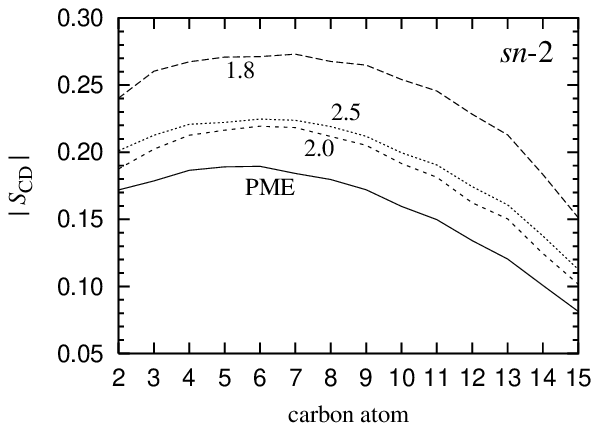,width=3.2in}

\caption{The order parameter $| S_{\mathrm{CD}} |$ 
         separately for the two acyl chains of DPPC:
         $sn$-1 chain and $sn$-2 chain. 
         For numbering of carbons, see 
         Fig.~\protect\ref{figDPPCmolecule}.}
\label{figOrderParameter}
\end{figure}

We have computed $S_{\mathrm{CD}}$ for all carbon atoms 
in both chains ($sn$-1 and $sn$-2) by averaging over all 
equivalent atoms in all DPPC molecules. The results are 
shown in Fig.~\ref{figOrderParameter}. We find that 
PME yields an order parameter profile which is in good 
agreement with experimental data \cite{Bro79,Dou95,Pet00}. 
Note that $S_{\mathrm{CD}} \approx 0.20$ close to the 
glycerol group, and tends towards zero toward the end 
of a tail. The acyl chains are therefore reasonably 
ordered close to the head group, while conformational 
disorder becomes more and more apparent towards
the center of the bilayer.

The results with different cutoff distances differ 
significantly from those obtained using PME. The 
situation close to the glycerol group clearly demonstrates 
the problem. The truncation of electrostatic forces at 
$r_{\rm cut} = 2.0$\,nm and 2.5\,nm yields order parameters 
that deviate about 13\% from the values obtained using PME. 
In the case of $r_{\rm cut} = 1.8$\,nm, the deviation is 
even larger, being of the order of 40\% close to the 
glycerol group. Differences are expected since the ordering 
of acyl chains must be affected by the packing of lipids as 
discussed in the previous section. A reduction in the area 
per molecule leads to an enhanced ordering of the acyl chains 
and correlates with an increased thickness of the bilayer.

\begin{figure}
\centering
\epsfig{file=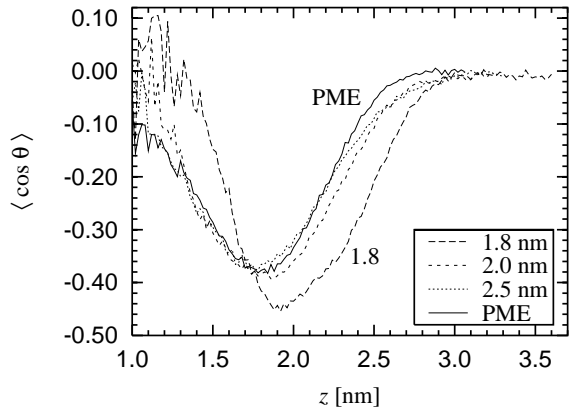,width=3.2in}
~\vspace{0.1cm}
\caption{$P(z) = \langle \cos \theta \, \rangle $ describing the ordering 
         of water close to the bilayer-water interface. $z = 0$ 
         corresponds to the center of the bilayer. The noise at 
         small $z$ is due to the small number of water molecules 
         inside the bilayer.}
\label{figWaterOrdering}
\end{figure}

We have also considered the ordering of water in the vicinity 
of the bilayer-water interface by calculating the time averaged 
projection of the water dipole unit vector onto the interfacial 
normal. As revealed by Fig.~\ref{figWaterOrdering}, the water 
molecules prefer to order themselves in such a way that the 
dipoles are oriented towards the bilayer. Ordering persists 
up to the height where the density of the lipids approaches 
zero.

This ordering  can be explained by the orientational behavior of 
water molecules around phosphoryl groups. The radial distribution 
of the oxygens and hydrogens of the water molecules around 
the phosphatidyl and choline groups were determined (data not 
shown), revealing that the water molecules are strongly 
oriented around the phosphorus atom. This is due to the hydrogen 
bonding between the oxygen atoms of the phosphatidyl group 
and the hydrogen atoms of the water molecules. Only a weak 
orientation is found around the nitrogen atom, most likely 
due to the hydrophobic nature of the surrounding methyl groups, 
which results in hydrogen bonding among the water molecules. 
This behavior is very similar for the different treatments of 
electrostatic interactions and also to earlier MD simulation 
studies of phosphatidylcholine systems~\cite{Alp93a,Dam94,Ess95b,Hyv97}. 
As the number of hydrogen-bonded water molecules around the 
phosphatidyl group is larger on the side of the water region, 
the dominating orientation of the water dipoles is toward the 
bilayer center. Only a few water molecules penetrate into the 
hydrocarbon region, resulting in poor statistics and a spiky 
profile.

To conclude this section, our results indicate that the ordering 
of fatty acyl chains is strongly affected by the method by which 
electrostatic interactions are treated. In addition, we find that 
the use of a relatively large cutoff in electrostatic interactions 
does not give rise to major artifacts in the properties of water 
molecules themselves or their radial distribution around the head 
groups. At the surface of the bilayer, however, packing of lipids 
affects the properties of the interfacial water layer. These 
conclusions support the view of previous research on the 
properties of water close to a water-lipid monolayer interface 
\cite{Alp93a,Alp93b,Fel96}. In these studies it was found that 
the artifacts were reduced by an increase in the cutoff, 
but were not eliminated for cutoff distances as large as 
$r_{\rm cut} = 1.8$\,nm. The work by Feller et al. provides 
a particularly interesting example of this issue, since they 
considered the radial distribution of oxygen-oxygen pairs in 
bulk water and found minor peaks close to the cutoff distance 
\cite{Fel96}. This is consistent with our results for RDFs close 
to the water-bilayer interface. We studied the radial distribution 
functions for O\,--\,N and O\,--\,P pairs (where O stands for 
oxygen in water) and found very weak but systematic ordering 
effects at $r_{\rm cut}$. In RDFs found by PME, such ordering 
effects were not present.

\subsection{Electrostatic potential}
\label{secElecProp}

\begin{figure}
\centering
\epsfig{file=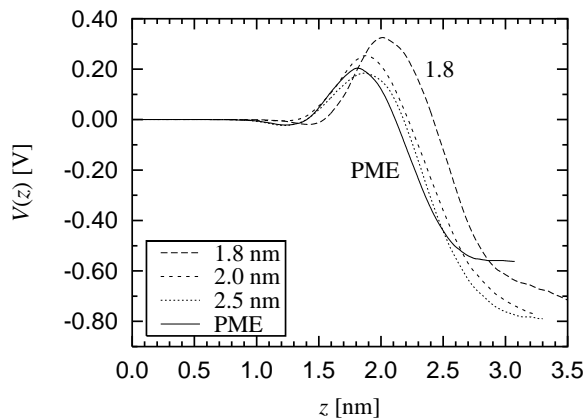,width=3.2in}
~\vspace{0.3cm}
\caption{Total electrostatic potential $V(z)$ across the bilayer, 
         where $z = 0$ corresponds to the center of the bilayer.}
\label{figPotentialBilayer}
\end{figure}

Based on the results discussed above, it seems obvious that 
the truncation of Coulombic interactions plays an important 
role in the electrostatic properties of the bilayer. To 
verify this and to quantify the magnitude of possible 
artifacts due to truncation, we studied the electrostatic 
potential $V(z)$. The results are shown in 
Fig.~\ref{figPotentialBilayer}.

The general behavior agrees with previous simulation 
studies of PC bilayers \cite{Tie96,Shi98b,Smo99,Smo00}. 
The lipid molecules contribute with a large positive potential, 
which is compensated by the contribution due to the water 
molecules. The total potential was determined to be $-570$\,mV.
For comparison, the experimental values for the potential 
range from $-200$\,mV to $-575$\,mV for different PC\,/\,water 
interfaces \cite{Fle86,Sim89,Gaw92,Mci92}.

~\vspace{-0.5cm}

As what comes to the differences between the curves in 
Fig.~\ref{figPotentialBilayer}, we note that the profiles 
of $V(z)$ are correlated with the mass density profiles in 
Fig.~\ref{figDensityBilayer}. The differences in the packing 
of lipids are  also reflected in the electrostatic potential: 
The strongest orientation of water molecules is found just 
above the peak in the density of lipids, and the orientation 
of the water ranges just as far as the head groups of the 
lipids prevail.

\subsection{Radial distribution of lipids}

We first consider RDFs between the two charged groups in 
a DPPC molecule. The positively charged choline group is 
at the top and the  negatively charged phosphate group 
at the lower part of the head group 
(see Fig.~\ref{figDPPCmolecule}). However, the average 
orientation of the P\,--\,N vector is almost parallel to the 
plane of the bilayer (data not shown). The details of the 
calculation are described in Sec.~\ref{secDataAnalysis}.

\begin{figure}
\centering

\epsfig{file=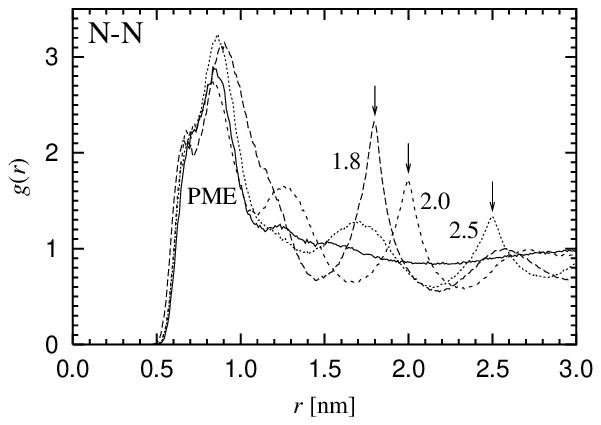,width=3.2in} 

\epsfig{file=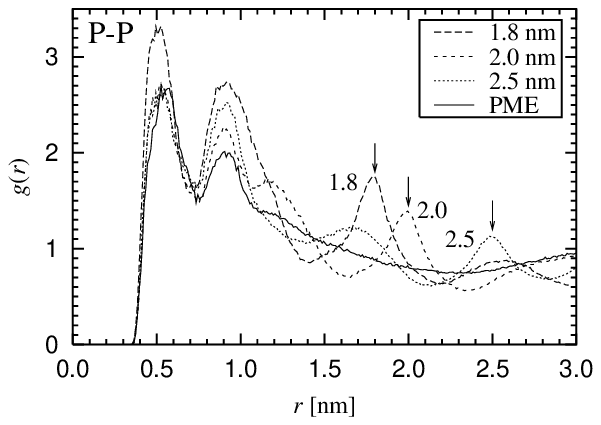,width=3.2in} 

~\vspace{-0.3cm}
\caption{Radial distribution function (RDF) $g(r)$ between the 
         two central atoms in the headgroup of a DPPC molecule: 
         RDFs for N\,--\,N and P\,--\,P pairs. Cutoff 
         distances are indicated by arrows.}
\label{figGroupRDF}
\end{figure}

The RDFs for the two pairs of P and N atoms in the head group 
are shown in Fig.~\ref{figGroupRDF}. The RDF of N\,--\,N pairs 
serves as a good example of our findings. The application 
of PME yields a radial distribution function which has 
a hard core at small distances, a rather narrow peak 
around 0.8\,nm, and essentially no structure beyond 
$r = 1.0$\,nm. This behavior is expected since we are 
dealing with a liquid-crystalline phase in the absence 
of translational order in the bilayer plane. The RDFs from 
the simulations in which a cutoff was used are dramatically 
different. In all of these cases we find that there is a wildly 
oscillating long-range component which has a local maximum 
exactly at the cutoff distance. In addition to this, the 
oscillations persist for distances far beyond $r_{\rm cut}$. 
Although the details are slightly different for P\,--\,P 
(as well as N\,--\,P) pairs, similar conclusions on artificial 
ordering can be drawn.

\begin{figure}
\centering
\epsfig{file=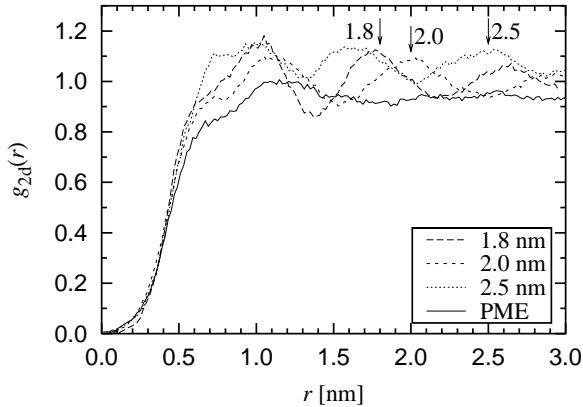,width=3.2in}
~\vspace{0.5cm}
\caption{Radial distribution function $g_{\rm 2d}(r)$ for 
         the center of mass positions of the DPPC molecules. 
         Note the relatively soft core of $g(r)$ in the 
         vicinity of $r = 0$. This is due to the fact that 
         the lipids may be entangled around each other such 
         that the $x,y$\,-coordinates of their center of mass 
         positions may lie close to each other. 
         }
\label{figCmRDF} 
\end{figure}

These structural artifacts have a very strong character. 
This is evident from the RDF for the DPPC center of mass 
positions presented in Fig.~\ref{figCmRDF}. The $g(r)$ given 
by the PME method is consistent with the assumption of 
a fluid-like phase, having just a small peak around 1.2\,nm. 
The truncation methods, on the other hand, give rise to further 
structure manifested as artificial maxima precisely at the 
cutoff distance.

Results of similar nature have been reported for ionic 
systems \cite{Gun98}. For an aqueous NaCl solution van Gunsteren 
and Mark found that the truncation of Coulombic interactions gave 
rise to an artificial peak around the cutoff distance. In our work 
the DPPC molecules are polar, though the behavior of the bilayer 
is not expected to be solely dictated by polar groups. In this 
sense, the pronounced artifacts in Fig.~\ref{figCmRDF} for lipid 
molecules may come as a surprise.

We can conclude that the truncation of electrostatic 
interactions gives rise to artificial order in the plane of 
the membrane. Truncation changes the phase behavior

\noindent
of the bilayers, and consequently affects 
thermodynamic properties such as the compressibility of the 
bilayer. It is clear that this is a matter of serious concern. 
It suggests that the truncation of Coulombic interactions in 
lipid bilayer systems may not only influence the short-range 
order of the system, but also the long-range behavior. 
Various intriguing phenomena such as organization 
of bilayer-protein systems involve scales of several molecular 
diameters. The artificial ordering observed in this work 
persists over these distances, and care should be taken 
when treating these systems.

\section{Summary and Conclusions}

We have investigated the effects of truncating Coulombic 
interactions on the properties of a DPPC bilayer and water 
in the vicinity of a bilayer-water interface. We have found 
that the truncation of electrostatic interactions can have 
serious consequences for the structural and electrostatic 
properties of lipid bilayer systems. The reduced area per 
molecule and the corresponding changes in the ordering of 
acyl chains provide two examples of the problems 
associated with truncation. As shown in this work, these 
artifacts are due to the fact that truncation gives rise 
to artificial ordering in the bilayer plane, which in turn 
implies that the lipid bilayer is no longer in a truly 
fluid-like state. Instead, the artificial order persists over 
a long range, which for typical system sizes studied by molecular 
dynamics simulations may exceed the actual size of the system. 
Furthermore, since the artificial order due to truncation affects 
the phase behavior of the system, we may conclude that 
there is indeed a reason for major concern.

The Particle-Mesh Ewald (PME) technique, on the other hand, 
has performed very well in this study. Besides providing results 
consistent with experimental data, it has given no reason for 
concern with respect to pair correlation behavior in the 
plane of the bilayer. The current trend of using PME in MD 
simulations of lipid membrane systems seems to be justified 
and should be encouraged.

We would like to stress, however, that even  PME and its 
variants may lead to artifacts unless great care is taken. 
These artifacts are related to the periodicity of 
the system, as periodic boundary conditions are used to 
eliminate finite size effects in simulations of small systems. 
H{\"u}nenberger et al.\ have observed \cite{Hun99,Web00} 
that the artificial periodicity used in Ewald techniques may 
indeed affect the conformational equilibria of e.g.\ peptides 
and proteins by stabilizing the most compact conformations 
of the molecules. It is clear that more attention is called 
for to develop more reliable and efficient techniques for 
dealing with electrostatic interactions in simulations of 
biomolecular systems.

\bigskip

{\em Acknowledgements} --- 
This work has, in part, been supported by the Academy 
of Finland through its Center of Excellence Program 
(E.F., P.L., and I.V.), the National Graduate School 
in Materials Physics (E.F.), the Academy of Finland 
Grant No.~54113 (M.K.), the Jenny and Antti Wihuri 
Foundation (M.H.) and the Federation of Finnish 
Insurance Companies (M.H.).


\vfill
\pagebreak

\appendix

\section{Effect of other simulation parameters}
\label{secAppendix}

We have complemented the main body of research 
by further simulations to check the importance of other 
simulation parameters in the present work. For example, 
we have checked the importance of constraints by keeping the DPPC 
bonds flexible, investigated the role of pressure coupling by a study 
in which the time constant of the barostat was set to 
a value of 5\,ps, and repeated all the simulations 
at a temperature of $T = 325$\,K. In all cases we have found 
that both the quantitative results and the conclusions remain
the same. However, changing the cutoff of the van der 
Waals interaction $r_{\rm vdW}$ appears to have effects  
worth noticing. While the parameterization 
of DPPC was done with a cutoff of $r_{\mathrm{vdW}} = 1.0$\,nm, 
and was therefore used in the major part of this study, 
the value of 1.4\,nm is also a common choice. This is motivated 
in particular by the Gromos\,96 \,forcefield which assumes 
$r_{\mathrm{vdW}} = 1.4$\,nm.

\begin{figure}
\centering
\epsfig{file=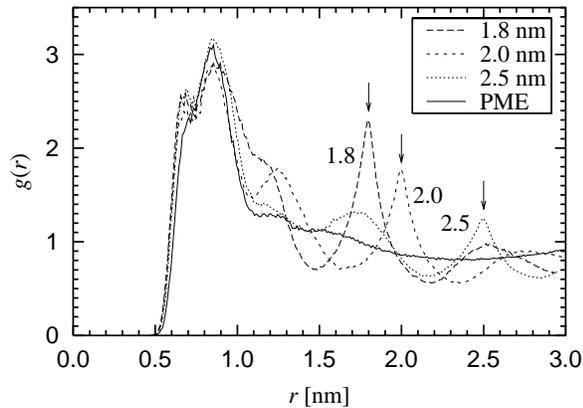,width=3.2in}
\caption{Radial distribution function for N\,--\,N pairs in 
         the head groups of DPPC molecules based on simulations 
         with a van der Waals cutoff $r_{\mathrm{vdW}} = 1.4$\,nm 
         [cf. Fig.~\protect\ref{figGroupRDF} with 
         $r_{\mathrm{vdW}} = 1.0$\,nm]. Cutoff distances 
         are indicated by arrows.}
\label{secRdfLJ}
\end{figure}

Simulations similar to those with $r_{\mathrm{vdW}} = 1.0$\,nm 
have been run with $r_{\mathrm{vdW}} = 1.4$\,nm. The average 
area per lipid using PME is reduced by 0.051\,nm$^2$, 
while the average areas per lipid for the three different 
cutoffs --- $r_{\rm cut} = 1.8$\,nm, 2.0\,nm, or 2.5\,nm --- 
are reduced by 0.017, 0.046, and 0.067\,nm$^2$, respectively. 
This trend is understandable since an increase in $r_{\rm vdW}$ 
effectively increases the attractive interaction between acyl 
chains, thus reducing the area per molecule. 
Interestingly, also the average volume per lipid is reduced 
by about 0.03\,nm$^3$, as was revealed by the Voronoi analysis 
of lipid bilayers in three dimensions. Despite these 
quantitative difference between the different cutoff distances 
for the van der Waals interactions, the conclusions of our work 
remain intact. This is demonstrated in Fig.~\ref{secRdfLJ}, 
which presents data for the RDFs between the N\,--\,N pairs 
with a cutoff $r_{\rm vdW} = 1.4$\,nm. The data indicate that 
the truncation of electrostatic interactions still gives rise 
to artificial ordering, which is not observed in the case of PME. 

\end{document}